\documentclass[11pt,twocolumn]{article}

\usepackage[round,numbers,sort&compress]{natbib}

\makeatletter
\renewcommand\@biblabel[1]{#1.}
\makeatother

\usepackage{amsmath,amssymb}
\usepackage{changebar}
\usepackage{times}
\usepackage{graphicx}
\usepackage[english]{babel}

\title{The role of the cytoskeleton in volume regulation and beading transitions in PC12 neurites}

\author{Pablo Fern\'andez$^\ast$ and Pramod A. Pullarkat$^\dagger$\\[2ex]
\\
\normalsize{$^\ast$ E27 Lehrstuhl f\"ur Zellbiophysik, Technische Universit\"at M\"unchen}\\
\normalsize{James Franck Stra{\ss}e, D-85748 Garching, Germany}\\
\normalsize{pfernand@ph.tum.de}\\[1ex]
\normalsize{$^\dagger$ Raman Research Institute}\\
\normalsize{C. V. Raman Avenue, Sadashivanagar, 560080 Bangalore, India}\\
\normalsize{pramod@rri.res.in}\\[2ex]
}
\date{\today}

\begin{document}

\maketitle

\abstract{
We present investigations on volume regulation and beading shape transitions in PC12 neurites conducted using a flow-chamber technique.
By disrupting the cell cytoskeleton with specific drugs
we investigate the role of its individual components
in the volume regulation response.
We find that microtubule disruption increases both
swelling rate and maximum volume attained,
but does not affect the ability of the neurite to recover its initial volume.
In addition, investigation of axonal beading --also known as pearling instability-- 
provides additional clues on the mechanical state of the neurite.
We conclude that the initial swelling phase
is mechanically slowed down by microtubules,
while the volume recovery is driven by passive diffusion of osmolites.
Our experiments provide a framework
to investigate the role of cytoskeletal mechanics
in volume homeostasis.\\[2ex]
}

\section*{INTRODUCTION}

The ability of living cells to regulate their volume is a ubiquitous
homeostatic feature in biology \citep{weiss,florian}. Since water readily permeates through the cellular membrane, alterations in extracellular osmolarity can change the concentration of all cytoskeletal components with severe consequences for
the metabolism.
Not surprisingly, one finds several mechanisms
involved in volume regulation.
In particular, many eukaryotic cell types display a short-term volume regulation
response to sudden alterations in external osmolarity, the
so-called regulatory volume decrease (RVD) and
regulatory volume increase (RVI) \citep{florian,uruguayos}.
It is generally accepted that these
require the passive diffusion of ions.
In the case of RVD, cell swelling upon hypoosmotic shock 
increases the membrane permeability for sodium,
which diffuses out of the cell. In turn, the decrease of intracellular osmolarity
drives water out and lowers cell volume \citep{florian,uruguayos}.
In this mechanism
water flow is driven by a difference in osmotic pressure.
It is often assumed that hydrostatic pressures
are negligible \citep{macknight}, with the argument that
they would make the membrane burst \citep{weiss}.
However, though the maximal pressures sustained by lipid bilayers
are indeed very low, the situation cannot be simply
extrapolated to the living cell.
The cell membrane is connected to the cytoskeleton,
the biopolymer gel spanning across the cell interior 
\citep{braybook,alberts,cytoskel_signal_janmey}. 
Since the cytoskeleton is viscoelastic and contractile \cite{glassy1,thoumine,mipaper,our_review}
(also in PC12 neurites \cite{pramod07_axon-mech}),
it may provide a mechanical memory of the initial state
as well as a driving force for volume relaxation
\citep{henson,kleinzeller,cantiello,mills,strieter,volume_strange,downey,pramods}.
Though it has been argued that the cytoskeleton is
too weak (typical moduli are up to 10 kPa \cite{glassy1,mipaper}) 
to sustain osmotic pressures (up to $\sim$MPa) \cite{cytoskel_signal_janmey},
neurons under hypoosmotic shock sustain strong pressures 
for several hours \citep{neuronsextreme} and
swelling of erythrocytes increases and approach perfect osmometer behaviour after disruption of the spectrin-actin cytoskeleton \citep{osmo_erythro_cytoskel_heubusch}.
This is indeed compelling evidence for a mechanical role, but in many systems the cytoskeleton seems to additionally play a signalling one.
In particular, biochemical interactions 
between actin filaments and ion channels
may couple strain of the actin cortex to changes in channel activity \citep{florian,suchyna}.
The fact that the actin cortex is disrupted when hypoosmotic
swelling begins \citep{d'alessandro,cornet88} seems to be
due to an influx of Ca$^{2+}$ \citep{Ca_swelling_Light}
through mechanosensitive ion channels activated by membrane stretching \citep{oscillations_Ca_stretch,Ca-waves_Ga}.
To clarify the role of the cytoskeleton
one must discern between pure mechanical and mechanosensing
responses, a difficult task requiring direct
measurements of membrane tension.

In this work, we study volume regulation in neurites,
axon-like cylindrical protrusions extended by PC12 cells \citep{PC12_establishment}, structurally very similar to the axons produced by neurons in culture \citep{alberts, changes_neurite_cytoskel_jacobs}.
Neurites furnish a hitherto unexplored, yet attractive model system 
to investigate the role of mechanical tension in volume regulation.
Their simple cylindrical geometry and low amount of
invaginated membrane allows better volume and area calculations
from images. They also have a well-defined, highly organized cytoskeletal structure
similar to that of axons: a central array of longitudinally arranged microtubules interconnected by microtubule binding proteins and surrounded by an actin cortex.
Moreover, an exceptional feature of axons and neurites is their ability to
undergo sudden shape transformations in response to an applied mechanical tension \cite{ochs1,ochs2,pramods}, an instability known as axonal beading to biologists and pearling instability to physicists. 
We show that axonal swelling increases membrane tension
and that microtubules slow down the volume response.
We conclude that frictional forces in the cytoskeleton
play an important role in axonal volume regulation.

\section*{EXPERIMENTAL SETUP AND METHODS}

\subsection*{Flow chamber}
\begin{figure}
\begin{center}
\includegraphics[width=0.45\textwidth]{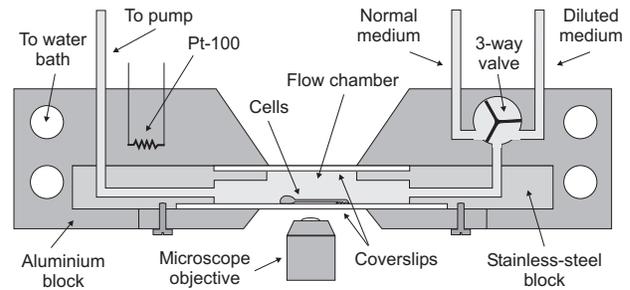}
\caption{\label{fig:fl-ch}\it\small
A schematic diagram of the flow-chamber set-up.
}
\end{center}
\end{figure}

The experiments have been carried out using a flow-chamber technique. 
A schematic of the set-up is shown in Fig.~\ref{fig:fl-ch}. A
stainless-steel block and two coverslips
are used to form a $10 \times 5 \times 1$ mm$^3$ chamber.
Cells adhere on the bottom cover-slip. One duct of the chamber
is connected to a peristaltic pump by means of long, soft silicone-rubber 
tubing that minimizes pressure fluctuations arising from the pump.
The other duct is connected to a 3-way-valve to select between two different 
media that can be pumped into the chamber. Stainless-steel
tubes that are $1$ mm in diameter connect the 3-way-valve to the two 
reservoirs where the media are stored. The chamber is intentionally made 
small to ensure a quick switching from one medium to the
other at  low flow rates, in the range of $2 - 4 \mu$l/s.
The switching of concentrations inside the chamber was studied
by adding an absorbing dye to one of the solutions and monitoring
the variation in the transmitted intensity with time.
The concentration reaches 90\% of its final value in about 10 s.
The chamber, the 3-way-valve, and the stainless-steel tubes are
placed inside an aluminium block with good thermal contact between each
other and a water bath maintains the temperature of the block at the
desired value. The continuous
flow of pre-warmed medium keeps the chamber at constant temperature
despite some heat loss through the cover-slips.

\subsection*{Cell culture}

PC12 cells are from the German Collection of Microorganisms and Cell Lines 
(DSMZ) \citep{DSMZ}. They are plated on collagen coated slides and cultured
in RPMI-1640 medium (Gibco) with 10\% fetal bovine serum and 5\% horse serum in presence of nerve growth factor (NGF) (Sigma-Aldrich Chemie, Munich, Germany) for 4--5 days \citep{PC12_culture_greene}. Such young neurites are known not to have intermediate filaments \citep{changes_neurite_cytoskel_jacobs}.
Slides are coated with 3-aminopropyl triethoxysilane (Sigma).
The silanised slides are covered with about 100 $\mu$l of
10\% rat-tail collagen (Sigma-Aldrich) dissolved
in a 70\% ethanol -- 30\% water solution
and let dry overnight.

\subsection*{Experimental procedure}
Prior to an experiment, 
the slide with the adherent cells
is transferred to the
flow-chamber. Cells are allowed to stabilize for
about 5 min by circulating the experiment medium
(normal medium with addition of 25 mM HEPES buffer (Invitrogen, Darmstadt, Germany)).
Experiments are performed by switching
from the normal experiment medium 
to experiment medium diluted with deionized water.
The response of the cells to the sudden switch in the external concentration is observed with an Axiovert 135 microscope (Zeiss, Oberkochen, Germany) configured for phase-contrast imaging.

\subsection*{Image analysis}
The volume and area of the neurites are analyzed from the recordings using
a home-made edge detection program. 
Edge detection using a threshold for intensity is unreliable due to the ``halo
effect'' present in phase-contrast images and also due to the dependence on the 
illumination intensity. To avoid such complications, edge pixels are recognized along the neurite by detecting the local maxima in the gradient of intensity across the neurite. After edge detection
the neurite volume and surface area are computed
assuming axial symmetry for the neurite
shape. Axial symmetry is verified by comparing the two detected boundaries
of several neurites and is found to be a good approximation for straight neurites
which are attached only at the two extremities. Only such
neurites were selected for the experiments.

\subsection*{Drug-induced cytoskeletal perturbation}
Experiments were performed in presence of cytoskeleton
disrupting drugs in order to study the role of its individual
components. A complication in these experiments arises
due to the neurites becoming fragile or losing their cylindrical
geometry on cytoskeletal perturbation. This precluded
experiments with the actin disrupting drug Latrunculin, which
induces detachment of the growth cone.
In the case of the microtubule disrupting drug Nocodazole (NOC) \citep{nocodazol}, which induces shape irregularities after approximately 10 min exposure, we let the drug act for 5 min before performing the hypoosmotic shock.
NOC concentration was 10 $\mu$g/ml throughout.
In contrast, the myosin-blocking drug Blebbistatin \cite{blebbistatin} did not significantly alter neurite shape. Thus, to ensure its effect we incubated neurites at $37^\circ$ for 1 hour at a high concentration (50 $\mu$M) before transferring them to the experiment chamber and performing the
experiment at a lower concentration of 20 $\mu$M.
For all experiments, drugs were present both in the normal and in the diluted medium. 
Since all drugs are dissolved in dimethylsulfoxide (DMSO),
which itself has effects on water and ion channels \citep{dmso_perm_hoek},
we performed control experiments in presence of DMSO 0.5\%, equal to the highest DMSO concentration in any of the drug experiments.

\section*{EXPERIMENTAL RESULTS}

\subsection*{Volume dynamics}

\begin{figure}
\begin{center}
\includegraphics[width=0.45\textwidth]{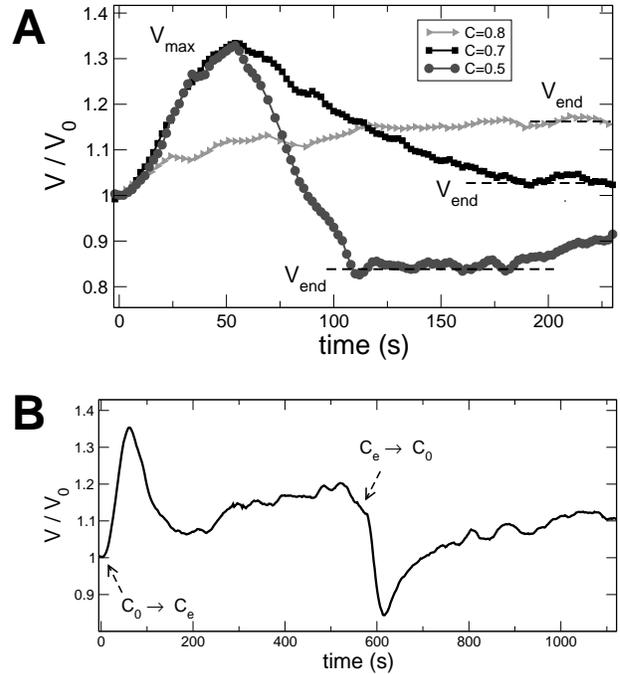}
\caption{\label{fig:osmoexp}\it\small
{\bf A:}
Swelling and recovery after a hypoosmotic shock performed at time $t=0$. Normalised volume $V/V_0$ as a function of time for different dilutions: 
$C=0.5$, ({\bf circles}),
$C=0.7$ ({\bf solid line}),
$C=0.8$ ({\bf dashed line}). Temperature is in all cases 36$^\circ$C. Each curve corresponds to a different neurite. The maximum volume attained is $V_{max}$, and the minimum is $V_{\rm end}$. Notice the strong undershoot of the volume at $C=0.5$, and the absence of recovery at $C=0.8$. These are all general trends of the volume response.
{\bf B:}
The full observed response when the neurite is subjected to a hypoosmotic shock and a subsequent hyperosmotic shock. At first the 
hypoosmotic shock 
$C_0 \rightarrow C_e$ is applied and the neurite is allowed to swell and relax (RVD). After the relaxation phase the hyperosmotic shock
$C_e \rightarrow C_0$ is applied and the neurite shrinks and recovers the initial volume (RVI). Such $C_0 \rightarrow C_e \rightarrow C_0$
cycles can be repeated up to five times in a given neurite before the ends detach (data not shown).
}
\end{center}
\end{figure}

We begin all experiments with a hypoosmotic shock imposed by
switching the cell culture medium flowing through the chamber from normal 
medium with a total solute concentration $C_0 \simeq 300$ mM to a lower 
external value $C_e$ (the continuous flow ensures constant external
concentration at all times).
In the following, for an intuitive measure of the shock strength we will
normalize the external osmolarity by its initial value and denote it by
$ C = C_e / C_0$.
Figure \ref{fig:osmoexp} shows typical responses for three different values of $C$ at
36$^\circ$C. For weak shocks ($C = 0.8$) the neurite volume increases from its initial volume $V_0$
until it reaches a maximum steady value $V_{max}$. No recovery is observed for tens of minutes. For intermediate shocks, $C = 0.7$, the volume increases at a roughly constant rate
$\dot{V}_0$ initially until it reaches a maximum value $V_{max}$.
Subsequently, the volume recovers almost back to $V_0$
with a typical regulatory volume decrease response (RVD).
The volume recovery is roughly exponential with a characteristic time $\tau$. The recovery time $\tau$ is strongly temperature-dependent (shown later),
but does not follow a clear trend with neurite diameter.
For strong shocks, $C = 0.5$, the recovery is
faster and there is often a remarkable ``undershoot'', where
the volume goes below its initial value.
Moreover, during the swelling phase the neurite undergoes
a sudden and transient shape transformation from its
normal cylindrical geometry to a periodically modulated one. As discussed later, this
Rayleigh-like ``pearling'' instability or beading indicates an increase in membrane tension upon neurite swelling. This shape transformation, which will be discussed later,
does not affect the volume response curve in any measurable way. This fact allow us to discuss
the volume dynamics and shape dynamics in that order.

Once the volume stabilizes 
to the lower external osmolarity $C_e$ (within about ten minutes), 
we perform a hyperosmotic shock by switching back the original medium
with concentration $C_0$.
The neurite shrinks, reaches a minimum volume  $V_{min}$ and then comes back to its initial volume in an RVI response
as shown in Fig.\ \ref{fig:osmoexp}.
No beading shape transformation is observed in this case.

In the following, we separately address the volume
regulation response and the peristaltic modulation. 
We study them as a function of osmotic shock strength,
temperature, and in the presence of cytoskeleton-disrupting drugs.

\begin{figure}[]
\begin{center}
\includegraphics[width=0.45\textwidth]{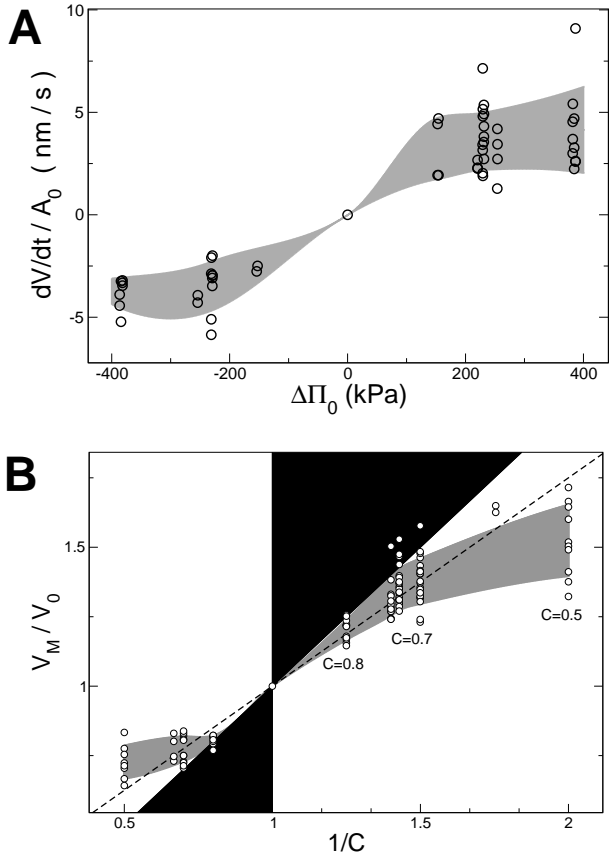}
\caption{\label{fig:rate-vmax}\it\small
{\bf A:}
Initial swelling speed $\dot{V}_0 / A_0$ as a function of the initial osmotic pressure difference $\Delta\Pi_0$ for temperatures $33-36^\circ C$. The shaded region is a guide to the eye, a spline going through the averages within a standard deviation. Notice the nonlinear response to strong hypo- as well as hyperosmotic shocks.
{\bf B:}
Maximum change in relative volume $V_M/V_0$ as a function of the 
inverse external concentration $1/C$, for all temperatures (here $V_M$ stands for $V_{max}$ or $V_{min}$ as the case may be).
The grey region is a guide to the eye.
The dashed line corresponds to a perfect osmometer
with dead volume $V_{\rm dead}=0.25 V_0$, and
the black regions correspond to negative dead volumes.
}
\end{center}
\end{figure}

{\bf Nonlinear swelling response.}
Figure \ref{fig:rate-vmax} A shows the initial rate of change of volume
divided by the initial area, $\dot{V}_0/A_0$,
as a function of the initial osmotic pressure difference
$\Delta\Pi_0 = RT C_0 (1-C)$,
for temperatures 33--36$^\circ$C.
Assuming zero hydrostatic pressure
and neglecting changes in internal osmolarity
gives the following expression for the initial swelling rate:
\begin{equation}
\label{eq:naive_linear}
\dot{V}_0 = A_0\,L_p \,\Delta\Pi_0 \;,
\end{equation}
where $L_p$ is the hydraulic permeability.
Surprisingly, from the average $\dot{V}_0$ value at $C=0.7$
we obtain an osmotic permeability $P_f=RT\delta_W\,L_p\simeq 1.4\,\mu$m/s
(where $\delta_W$ is the molar density of water),
which is about two orders of magnitude
lower than the literature values for
lipidic membranes and most biological cells
\citep{Wperm_renal_maric, Wperm_memb_huster, brasileros, weiss} even after blockage of water channels \citep{dmso_perm_hoek}.
Moreover, as can be seen from Fig.\ \ref{fig:rate-vmax} A, 
the expected linear dependence is contradicted by the strong nonlinear response
observed for both hypo- and hyperosmotic shocks.

We now turn to the maximum (minimum) volume attained in a hypoosmotic (hyperosmotic) shock.
The maximum (minimum) volume $V_{max} (V_{min})$ is to a good approximation proportional to the initial volume $V_0$ and does not depend significantly on the temperature (data not shown \cite{mitesis}). Thus we look at the relative maximum volume $V_{max}/V_0$ in order to minimize the effect of neurite diameter. As shown in Fig.~\ref{fig:rate-vmax} B,
the maximum volume increases nonlinearly 
with the initial osmotic pressure difference $\Delta\Pi_0$.
The data is contrasted to the perfect osmometer equation,
corresponding to a constant total amount of internal osmolites
and zero hydrostatic pressure:
\begin{equation}
\frac{V_0 - V_{\rm dead}}{ V_{max} - V_{\rm dead} } = C \;,
\end{equation}
where the dead volume $V_{\rm dead}$ represents non-aqueous internal volume. Mammalian cells have on the average a cytosolic protein concentration of $\sim 20\%$ \citep{braybook}. According to electron microscopy studies \citep{changes_neurite_cytoskel_jacobs} the non-cytosolic volume of PC12 neurite is comprised mostly of microtubules and organelles and amounts to $V_{\rm dead} \simeq 25\%\,V_0$. 
The black region in Fig.\ \ref{fig:rate-vmax} B 
corresponds to $V > V_0 / C$ for hypoosmotic shocks and to $V < V_0 / C$ for hyperosmotic shocks;
penetrating this region 
would require work against the osmotic gradient. 
As expected for a passive response, 
the data lies outside them.
For mild shocks, $C \geq 0.7$ and (hyperosmotic) $C \leq 1.4$, 
neurites swell as much as perfect osmometers 
a dead volume of 25\%. Therefore, any ion leakage
or hydrostatic pressure are negligible during the swelling phase.
However, at strong shocks (hypoosmotic $C=0.5$ or hyperosmotic $C=2$),
neurites swell significantly less than a perfect osmometer
would. Thus, upon strong osmotic perturbations
either ions leak or a sustained hydrostatic pressure
develops.


{\bf Volume regulation under cytoskeleton disruption.}
The axonal cytoskeleton may be expected to contribute to the volume 
response in several ways. As discussed in the introduction, a mechanical as well as a signalling role is conceivable.
In order to assess the role of individual components of the cytoskeleton,
we treat neurites with the myosin blocking 
drug Blebbistatin (BLE) \citep{blebbistatin}
and the microtubule disrupting drug
Nocodazole (NOC) \citep{nocodazol}.
Since these are all diluted in dimethylsulfoxide (DMSO),
a compound known to alter ion channels, we also
perform control experiments in presence of DMSO.

\begin{figure}[]
\begin{center}
\includegraphics[width=0.45\textwidth]{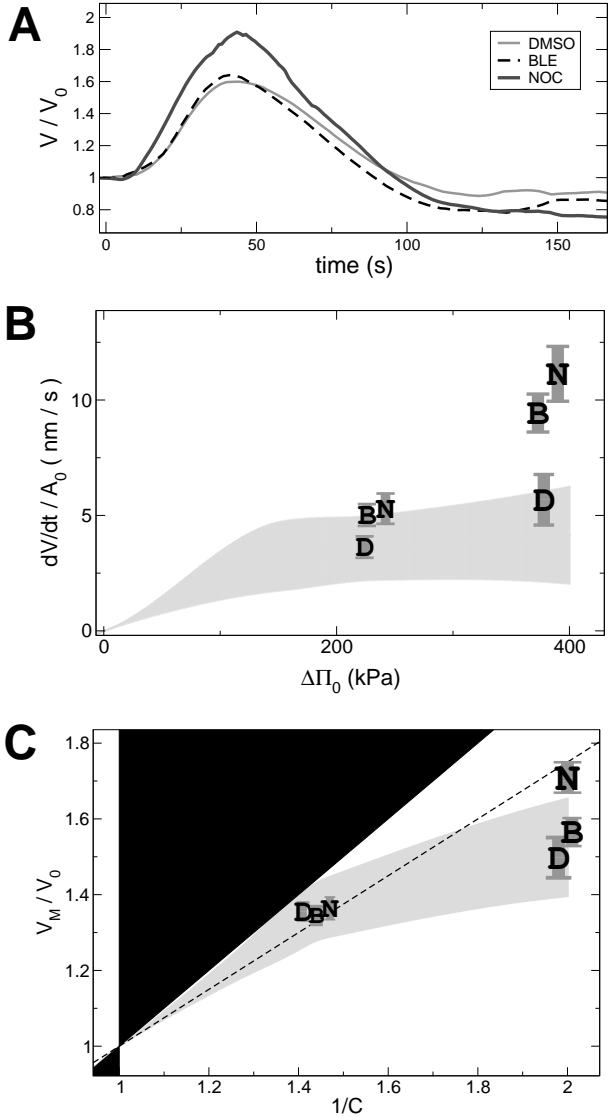}
\caption{
\label{fig:Swell_drugs}\it\small
{\bf A:}
Comparison of different drug treatments. Each curve is a different
neurite. All experiments were performed at temperature $T=33^\circ$C
and dilution $C=0.5$. Grey shaded line: DMSO control.
Dashed line: BLE.
Black solid line: NOC.
{\bf B:}
Swelling speed $\dot{V}_0 /A_0$ as a function of the initial osmotic pressure difference $\Delta\Pi_0$, for temperatures $33-36^\circ C$. 
{\emph D}: DMSO control ($n=5$ for $C=0.5$, $n=14$ for $C=0.7$).
{\emph B}: blebbistatin ($n=7$, $n=13$). 
{\emph N}: nocodazol ($n=8$, $n=14$).
The grey region is a guide to the eye, corresponding to the experiments without drugs shown in Fig.\ \ref{fig:rate-vmax} A. Data for all drugs is shown as arithmetic mean $\pm$S.~E.
{\bf C:}
Maximum relative volume $V_M/V_0$ as a function of the osmotic pressure difference $\Delta\Pi$
in presence of cytoskeleton-disrupting drugs. 
The grey region 
corresponds to the data without drugs in Fig.\ \ref{fig:rate-vmax} B.
}
\end{center}
\end{figure}

Fig.\ \ref{fig:Swell_drugs} A shows typical responses.
In presence of Nocodazole
the initial swelling rate for strong shocks $C=0.5$,
increases markedly, but it barely changes for $C=0.7$
(see Fig.\ \ref{fig:Swell_drugs} B).
With Blebbistatin we observe
a weaker but still significant effect.
For both drugs, the relationship between
swelling rate and initial osmotic pressure difference
approaches the naively expected linear dependence
given by Eq.\ \ref{eq:naive_linear}.

Nocodazole induced disruption of microtubules
also has a strong effect on the
maximum volume $V_{max}$ attained after a strong shock of $C=0.5$,
as shown in Fig.\ \ref{fig:Swell_drugs} C.
Neither BLE-treatment nor DMSO alone 
have a significant effect on the maximum volume.
For mild dilutions, $C=0.7$,
nocodazole has no effect, consistent
with the fact that neurites swell like 
a perfect osmometer with 25\% dead volume.

Importantly, the cytoskeleton disrupting drugs
do not affect the ability of the neurite to perform RVD.
As can be seen in Fig.\ \ref{fig:Swell_drugs} A, 
the volume fully relaxes back to its initial value.
Further evidence can be found in the Supplementary
Material.

\begin{figure}
\begin{center}
\includegraphics[width=0.45\textwidth]{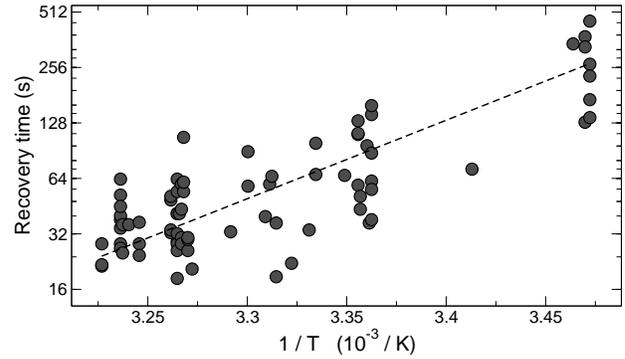}
\caption{
\label{tau_T}\it\small
Recovery time as a function of inverse temperature 1/T. The dashed line is a least-squares fit to an Arrhenius form 
$\tau\propto e^{\Delta G\,/\,RT}$.
}
\end{center}
\end{figure}

{\bf The effect of temperature: Arrhenius behaviour.}
We now address the influence of temperature in
the dynamics of volume regulation.
For simplicity, we describe the recovery phase
by fitting single exponentials.
As shown in Fig.\ \ref{tau_T},
lowering the temperature from 35 to 15$^\circ$C
slows down the volume dynamics by an order of magnitude.
The dashed line is an Arrhenius-like equation $\tau \propto 1/k \propto e^{\frac{\Delta G}{RT} }$, where $k$ depends on 
temperature as the rate constant of a thermally-activated process. 
This yields an activation energy
$\Delta G  \sim$ 30 kT, a typical order of magnitude
for biological processes \citep{biophysics_blumenfeld}.
The well-defined Arrhenius trend is consistent with the idea that
ion channels are responsible for RVD.

\subsection*{Beading}

\begin{figure}[]
\begin{center}
\includegraphics[width=0.45\textwidth]{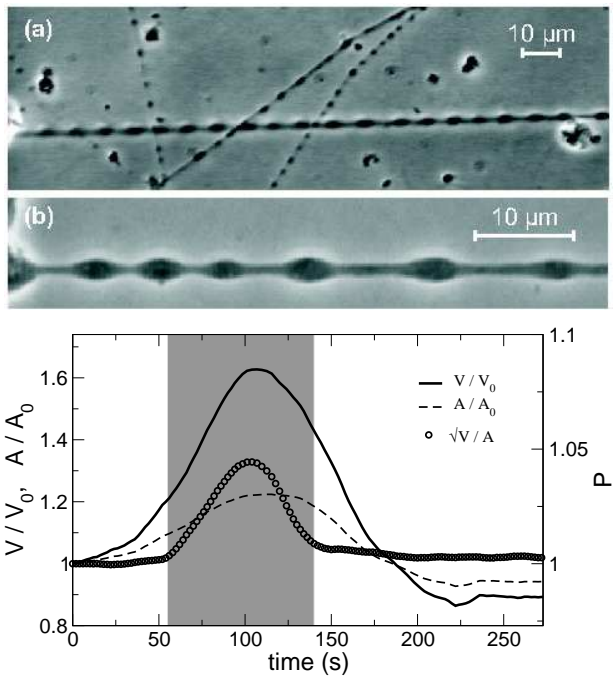}\\[1ex]
\includegraphics[angle=0,width=0.45\textwidth]{Fig6b.eps}
\caption{\label{fig:pearl}\it\small
{\bf Top:} 
Swelling-induced pearling instability. {\bf a:} Chick-embryo neurons.
{\bf b:} PC12 neurites at a higher magnification.
{\bf Bottom:}
Pearling instability. Relative volume $V/V_0$, area $A/A_0$ and shape
parameter $\sqrt{V/V_0}/(A/A_0)$ as a function of time.
Imposing a fast hypoosmotic shock at dilution $C=0.5$ and
temperature 33$^\circ$C
increases membrane tension and leads to a shape instability.
The instability vanishes before the volume recovers.
}
\end{center}
\end{figure}

When neurites are subjected to strong hypoosmotic shocks they undergo a shape transformation by developing a periodic array of swellings akin to beading of axons
\citep{08barbee_beading}. We observe this in chick dorsal root ganglia (DRG) neurons as well as PC12 neurites Fig.\ \ref{fig:pearl}. This peristaltic deformation, 
resembles that formed in  nerves subjected to induced stretch injuries \cite{ochs1,ochs2,ochs4} and that observed after traumatic injuries to the brain 
\cite{08barbee_beading}.
The dynamics of bead formation and the mechanism has been investigated recently using the osmotic shock technique \citep{pramods} and shown to be similar to the pearling instability observed in synthetic membrane tubes under tension  \citep{barziv1,barziv3}.
Here, we provide direct evidence correlating neurite tension and bead formation and
describe the role of cytoskeletal components in this process.

We begin by listing the main features of osmotic shock induced beading.
For a given radius of the neurite, there is a critical hypoosmotic shock below
which the shape remains cylindrical during the entire volume evolution, and above which a transient peristaltic modulation is observed. This is about 
$C=0.5$ at 37$^\circ$C and about $C=0.7$ at 25$^\circ$C for an initial neurite radius of 0.7 $\mu$m. 
Beading and recovery cycles (for mild shocks) can be repeated up to five times in the same neurite, after which neurites tend to detach from the substrate. Also transport of organelles can be observed during and after beading and recovery. 
This suggests that mild osmotic shocks leading to beading and recovery causes no permanent damage to the neurites.
On the contrary, strong shocks ($C \ge 0.5$) often lead to strongly modulated beaded shapes with no recovery for several
tens of minutes. Extreme shocks cause neurites to burst with organelles spewing out, corroborating a build up of membrane tension.
Finally, no beading is observed during hyperosmotic shocks, irrespective of the magnitude.

\begin{figure}
\begin{center}
\includegraphics[width=0.45\textwidth]{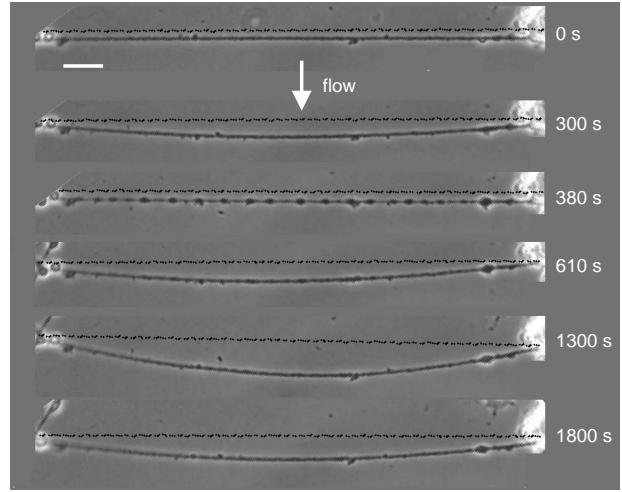}
\caption{\label{fig:pearl-flow}\it\small
Neurite deformations induced by the combined effect of a constant laminar flow perpendicular to the neurite
and different osmotic conditions. The temperature is $25^\circ$C and $C_e = 0.7 C_0$. The sequence of events is as follows: laminar 
flow is started at t = 20 s and the
neurite allowed to undergo viscoelastic relaxation and attain a steady state, 
a hypoosmotic shock is applied at t = 320 s by introducing dilute medium, the neurite is allowed to
undergo volume and shape relaxation, normal medium is reintroduced after the volume has relaxed
(hyperosmotic shock) at t = 1200 s, the neurite is again allowed to undergo volume relaxation. The flow rate is 
constant throughout the experiment.
The dashed lines are for comparison of curvature.
The larger the average  curvature of the catenary, the smaller the tension in the neurite \citep{pramod09_drag}.
}
\end{center}
\end{figure}

{\bf Axonal tension causes beading.}
The flow chamber technique has been used recently to quantitatively study the evolution of neurite tension in response
to drag forces and to demonstrate active contractile responses in neurites  \citep{pramod09_drag}.
This is performed by imposing a constant, laminar flow perpendicular to the neurite generating a drag force. 
It has been shown that PC12 neurites as well as axons are viscoelastic \citep{pramod07_axon-mech} and respond 
to a stretching force via a relaxation process with a characteristic timescale. Under the influence of a flow induced drag 
force the neurite takes the form of a catenary and elongates until it reaches a final equilibrium shape \citep{pramod09_drag}
(see Fig.\ \ref{fig:pearl-flow} t=300s). The resulting strain gives a measure of the tension in the axon.
Fig.\ \ref{fig:pearl-flow} shows snapshots of the neurite taken at various stages during an osmotic shock experiment.
Under the influence of the flow the neurite attains an equilibrium catenary shape within about a minute.
When the flow is switched from the normal medium to a diluted one, keeping the flow rate the same, the neurite shortens until
it becomes almost straight. The beading sets in during this straightening phase and  the maximum beading amplitude occurs when
the neurite is straight (t=380s in Fig.\ \ref{fig:pearl-flow}). Subsequently there is a relaxation process in which the beading amplitude begins to decay and
the flow-induced curvature increases. When a hyperosmotic shock is applied after the first volume
relaxation process (t=1200s in Fig.\ \ref{fig:pearl-flow}) the neurite curvature begins to increase  as the neurite shrinks and subsequently decreases
to almost its normal value with a timescale very much comparable to the volume recovery time. These observations clearly demonstrates that the neurite tension
and the volume changes induced by osmotic treatments are correlated.

{\bf Beading mechanism.}
The physical mechanism for bead formation is as follows \citep{pramods}. After applying 
an osmotic shock the
neurite volume increases as discussed above. This results in a corresponding 
expansion of the measured area $A(t)$ as shown in  Fig.\ \ref{fig:pearl}. This stretching of
the membrane causes an increase in the tension of the outer membrane and hence
an increase in the surface energy, $F_s = \sigma(A(t) - A_0)^2/A_0$. The increase in volume
also costs bulk elastic energy due to the deformations caused in the cytoskeletal
network, which we will consider below later.

It can be shown \cite{barziv3} that at
any instant $t$, for a volume $V(t)$, a peristaltically modulated shape
with 
\[
r(t, z) = r_0 + \epsilon (t) \sin(qz)
\]
has a lower average surface area 
compared to a cylinder with the same instantaneous volume $V(t)$, provided the
wavelength $\lambda > 2\pi r_0$. For small amplitudes, the relative area gain can be obtained as
\[
\delta S/S = \epsilon (\tilde{q}^2-1) / (4 r^2) \; ,
\]
where S is the surface area and 
$\tilde{q} = 2\pi r/\lambda$. This can be seen in Fig.\ \ref{fig:pearl}. The
ratio $\sqrt{V(t)}/A(t)$ (note: the average volume and area are computed for unit length)
increases when the peristaltic mode grows. This quantity 
is a constant for all cylinders irrespective of radius. An increase in  $\sqrt{V(t)}/A(t)$
from this constant value indicates a decrease in area for the peristaltic shape as 
compared to a cylinder with identical volume. In other words, the neurite is able to reduce
its interfacial energy by adopting a peristaltic shape instead of a cylindrical one.
As mentioned earlier any deformation costs bulk elastic energy due to the 
cytoskeletal elasticity. Unlike the surface free energy, the bulk elastic free energy is 
always positive for peristaltic modes compared to 
a cylinder of same volume (it costs bulk energy to expand as well as compress regions 
along the neurite).
Therefore, there is a competition between the surface energy and bulk energy which
determines the preferred shape, giving rise to a critical tension. 
A rough expression for the critical tension can be obtained  
by comparing the average bulk energy per unit length of a cylinder $E r_c ^{2}$, 
where E is the elastic modulus of
the cytoskeleton, and the corresponding surface energy $\sigma r_c $. This gives
a critical tension $\sigma_c = E r_c$, above which the surface contribution dominates. 
As seen in experiments, the critical tension needed to
destabilise the cylinder increases with neurite radius.  
Analogous shape transformations observed
in membrane tubes, triggered by application of laser tweezers \citep{barziv1,barziv3},
and in cell protrusions after treatment with latrunculins \citep{barziv2} have been 
analysed in a similar fashion.

The above mentioned expression for relative gain in area shows that reduction in area
increases with wavelength of the perturbation, giving a maximum area gain for $\lambda = L/2$,
$L$ the length of the neurite. Clearly, the observed value of   $\lambda$ is much shorter than and 
independent of the neurite length, but increases linearly with radius. A simple minded argument for this
is as follows. Any perturbation with a wavelength $\lambda$ generates as pressure difference given by the Laplace law, 
the pressure difference $p$ balanced by the membrane tension $\sigma$.
In the case of the axisymmetric modes there are two principal curvatures: that with a curvature radius equal to
the neurite radius $r(z)$, and that corresponding to the peristaltic deformation along the 
$z$-axis, with a curvature radius $1/\partial^2_zr(z)$. 
For large wavelengths, $\lambda \gg \epsilon$, the latter can be ignored.  The pressure difference in between 
the crest and the trough of a peristaltic mode can be written in the small amplitude limit ($\epsilon \ll r$) as 
\[ \Delta p = \sigma (1/r_{crest}- 1/r_{trough}) \simeq - \sigma \epsilon / r^2 \;.\]
As indicated by the negative sign, this is an unstable flow: the pressure difference will drive water from the troughs 
into the crests, thereby increasing the perturbation amplitude $\epsilon$. The flow rate and hence the growth of a given 
mode is in general proportional to the driving force
\[\Delta p / \lambda \sim \sigma\epsilon\, / \,\lambda r^2 \;.\] 
Shorter wavelength modes are thus faster to grow than longer wavelength ones.
However, for very short wavelengths the curvature along the neurite axis becomes important. The respective pressure 
difference is given by $\Delta p = \sigma \partial^2_zr(z) \simeq \sigma \epsilon / \lambda^2$.
In contrast to the previous case, here the cylindrical state is stable: water flows from the crests into the troughs 
(which have a ``negative'' pressure) making the perturbation vanish. Moreover, the dependence of the flow rate on 
the wavelength is more pronounced,
\[ \Delta p / \lambda \sim \sigma \epsilon / \lambda^3 \;.\] 
Therefore for very small 
wavelengths the stabilizing flow driven by the curvature along $z$ always dominates, and there is no instability anymore.
We conclude that there is a fastest mode at an intermediate, non-zero wavelength. This can be shown to be $\lambda \simeq 9.2 \,r$ \citep{pearling1}.
In the very dawn of linear stability analysis, this argument was applied to the case of water jets by Lord Rayleigh \cite{rayleigh, chandrasekhar}.

\section*{THEORY}

We model volume regulation in neurites
taking into account both mechanical and osmotic driving forces.
The neurite is characterised by its volume, $V(t)$,
and the internal amounts of ionic species, $n_i(t)$.
For simplicity we consider only potassium and chloride,
by large the most important osmolites inside the cell.
Since due to electroneutrality
the flows of Cl$^-$ and K$^+$ are coupled,
\[\dot{n}_{\rm Cl}=\dot{n}_{\rm K}\;,\]
one of the two concentrations can
be eliminated. The two variables of our model are
the adimensional quantities
\begin{eqnarray}
\mathcal{V}&=&V(t)\,/\,V_0 \\
\mathcal{N}&=&n_{\rm Cl}(t)\,/\,n_{\rm Cl}(0)\;.
\end{eqnarray}

The flow of water through the membrane \cite{weiss} is given by
\begin{equation}
\dot{V}=A L_p (\Delta\Pi - \Delta p)\;,
\end{equation} 
where $A$ is the neurite area,
$L_p$ the hydraulic permeability,
$\Delta p$ the hydrostatic pressure,
and the osmotic pressure difference is given by
\[\Delta\Pi \simeq RT \sum_i (n_i/V - c_i^{\rm ex})\]
in terms of external $c_i^{\rm ex}$
and internal osmolite concentrations $n_i/V$.
The typical relaxation time for the volume is
\[\tau_V= \frac{V_0}{A L_p \Pi_0}\;,\] 
where $\Pi_0 = RTC_0 \simeq 700$ kPa the osmotic pressure of normal medium.
Taking $10^{-14}$ m Pa$^{-1}$ s$^{-1}$ for the permeability
(from Fig.\ \ref{fig:rate-vmax} A) we obtain $\tau_V = 50$ s.

We model ion movement
with a passive K$^+$-Cl$^-$ cotransport
following Ref.\ \cite{uruguayos},
neglecting the effect of ion pumps since
these are not relevant for the short-term
RVD response.
Chloride flux is given by the
difference in chemical potentials,
\begin{equation}
\dot{n}_{\rm Cl} = -ARTG \log\bigg(
\frac{n_{\rm Cl} n_{\rm K}}{V^2}
\frac{1}{C^2 c_{\rm Cl}c_{\rm K}}
\bigg)\;.
\end{equation}

We model the RVD response following the standard
assumption of a volume-dependent permeability \cite{florian}.
The permeability $G$ must be 
zero at the initial volume $V_0$ and non-zero 
at a significant departure.
We decompose \[G(\mathcal{V})=\lvert G\rvert\; g(\mathcal{V})\;,\] 
into a typical order of magnitude $\lvert G\rvert$
and an adimensional function $g(\mathcal{V})$.
For the chloride relaxation time 
\[\tau_{Cl} = \frac{V_0 \: c^{\rm in}_{Cl}}{ART\lvert G\rvert}\;,\]
we expect a value of 10--100 s taking typical values for ion conductivities \citep{weiss}.
The volume-dependent permeability function $g(\mathcal{V})$
jumps from zero to a finite value at a critical threshold volume;
otherwise its form is not known and different expressions
have been used in the literature \citep{uruguayos,strieter}.
In our experiments the critical volume is about $1.2\,V_0$,
as can be seen in Fig.\ \ref{fig:osmoexp}.
We obtain good results with the simple expression
\[ g = \theta(\mathcal{V}/1.2) \: \mathcal{V} \;,\]
where $\theta()$ is the step function.

Finally, assuming external ion concentrations to jump instantaneously
at time $t=0$
from $c_i^{\rm ex}(0)$  to $C c_i^{\rm ex}(0)$
the full system can be written as
\begin{eqnarray}
\dot{\mathcal{V}} &=& \frac{1}{\tau_V} 
\left(
\frac{1+2 \phi(\mathcal{N}-1)}{\mathcal{V}} 
- \frac{\Delta p}{\Pi^0}  - C
\right)  \\[1ex]
\dot{\mathcal{N}} &=& - \frac{g(\mathcal{V})} {\tau_{Cl} } 
\log \left(
\;\frac{\alpha \; \mathcal{N}^{\,2}  +  (1-\alpha) \; \mathcal{N} }
{\beta C^2 \mathcal{V}^2 }
\;\right) 
\end{eqnarray}
with the parameters
\begin{eqnarray}
\phi &=& c^{\rm in}_{\rm Cl}\,/\,C_0\\[1ex]
\alpha &=& c^{\rm in}_{\rm Cl}\,/\,c^{\rm in}_{\rm K}\\[1ex]
\beta &=& c^{\rm ex}_{\rm Cl} \,c^{\rm ex}_{\rm K} \; / \;
c^{\rm in}_{\rm Cl} \, c^{\rm in}_{\rm K}  \;.
\end{eqnarray}
Taking the following physiological values for the ion concentrations \cite{weiss}
\begin{eqnarray*}
c^{\rm in}_{\rm Cl}&=& 80\,\mbox{mM} \\
c^{\rm in}_{\rm K} &=& 150\,\mbox{mM} \\
c^{\rm in}_{\rm Na} &=& 15\,\mbox{mM}  \\
c_{\rm Cl}^{\rm ex}&=& 80\,\mbox{mM} \\
c_{\rm K}^{\rm ex} &=& 6\,\mbox{mM} \\
c_{\rm Na}^{\rm ex}&=& 110\,\mbox{mM} \;,
\end{eqnarray*}
the adimensional constants become $\phi=0.25$,
$\beta \simeq 0.05$, and $\alpha \simeq 0.5$.

\begin{figure}
\begin{center}
\includegraphics[width=0.45\textwidth]{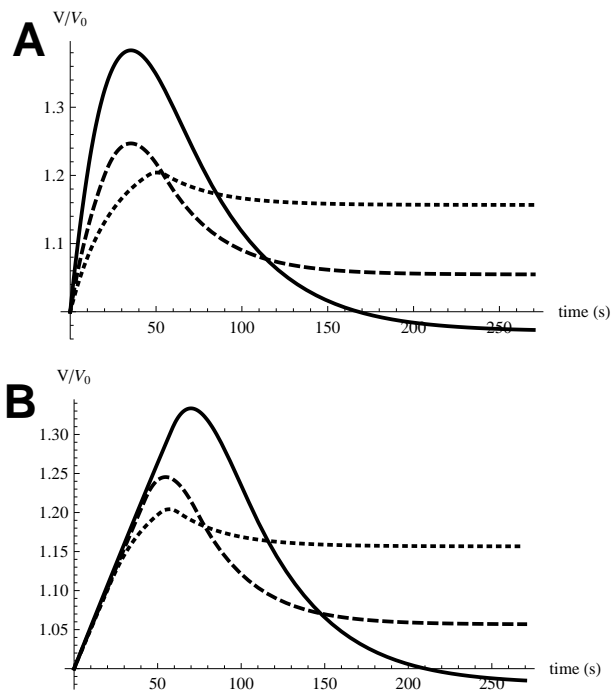}
\caption{\label{fig:model}\it\small
{\bf A:} zero pressure model for dilutions 80\% (dotted line), 70\% (dashed line) and 50\%
(solid line).
{\bf B:} adding nonlinear friction as given by Eq.\ \ref{eq:flow} reproduces the trend during the swelling phase.
Compare to Figs.\ \ref{fig:osmoexp},\ref{fig:pearl}
}
\end{center}
\end{figure}

{\bf Zero pressure model.}
The standard assumption in modelling volume
regulation is to set the hydrostatic pressure to zero \cite{uruguayos},
\[
\Delta p=0 \;.
\]
With this ansatz we can solve the equations
for the volume response $\mathcal{V}(t)$.
As shown in Fig.\ \ref{fig:model},
the model reproduces the timescales for swelling
and recovery as well as the maximum volume.
Interestingly it reproduces the ``undershoot'' at $C=0.5$
(see Fig.\ \ref{fig:osmoexp}).
However, the dependence of the
swelling rate on osmotic pressure is clearly off.
The model gives a linear response --a feature intrinsic to any volume-dependent permeability-- whereas experiments show a nonlinear one (Fig.\ \ref{fig:rate-vmax} A).

{\bf Viscoelastic model.}
As shown by our experiments combining
hypoosmotic shocks and drag forces (Fig.\ \ref{fig:pearl}),
neurites develop tension while swelling.
This suggests that the nonlinear response
observed in experiments may come from
a nonlinear mechanical response,
a ubiquitous feature of cells with
structured cytoskeleton \cite{fernandez2}.
We therefore assume that the pressure difference
$\Delta p$ is not zero, but depends on
the swelling rate as expected for
a viscoelastic element.
To reproduce Fig.\ \ref{fig:rate-vmax} A,
we assume that neurites swell at
zero pressure below a critical rate but
encounter internal friction for rates larger than
a critical value $\Omega$.
This can be written as
 \begin{equation}
 \label{eq:flow}
\Delta p = \left\{
\begin{array}{l@{\quad \mbox{for} \quad}l}
0 & \lvert\dot{\mathcal{V}}\rvert < \Omega \\
\eta \:{\rm sign}(\dot{\mathcal{V}})\: 
\big( \lvert\dot{\mathcal{V}}\rvert / \Omega - 1 \big) 
& \lvert\dot{\mathcal{V}}\rvert > \Omega
\end{array} \right.
\end{equation}
where $\Omega$ is the maximal rate of change of volume
at which the neurite can swell without tension, corresponding
to about 5 $10^{-3}\,{\rm s}^{-1}$ in our experiments,
and $\eta$ is the friction scale, about 2 MPa.

As shown in Fig.\ \ref{fig:model},
with this ansatz the model reproduces the essentially
constant swelling rate observed in experiments.
The nonlinear friction slows down both swelling and recovery.
Interestingly, the peculiar triangular shape of
of the curves at $C=0.5$ is much closer to that of experiments
(compare to Fig.\ \ref{fig:osmoexp}).

\section*{DISCUSSION}

{\bf Microtubules mechanically slow down swelling.} 
The initial response of neurites to
hypo- as well as hyperosmotic shocks
is a strongly nonlinear
function of the external osmotic pressure.
Remarkably, this response becomes much
simpler after microtubule disruption:
neurites behave as perfect-osmometers,
and the swelling rate increases linearly with osmotic shock strength.
The simplest explanation is that
microtubules slow down the initial volume
change by mechanically opposing the
hydrostatic pressure difference.
This is not exactly an elastic response 
but rather a (nonlinear) viscous one
which depends crucially on the swelling rate.
This mechanical role of microtubules may
come as a surprise, as they are in general
irrelevant for the (passive) mechanical response of
cells. In neurites, however, their
structure is different: they are arranged in
bundles interconnected by microtubule binding
proteins (MBPs). Our results indicate
that microtubules are firmly connected
to the membrane, either directly by MBPs
or through the actin cortex.
Interestingly, studies of volume regulation on round PC12 cells 
without neurites give a different picture.
Whereas disruption of the actin cytoskeleton by Cytochalasin B
has been reported to increase KCl efflux and diminish extent
of swelling \citep{cornet88,cornet93},
disruption of microtubules has no effect on RVD \citep{cornet88}.
This is consistent with our conclusion of a mechanical role for microtubuli,
since these are organised very differently in round cells and neurites:
in the latter their bundle structure provides a rigid scaffold
which can oppose swelling.

The mechanical response of the neurite, as
described by Eq.\ \ref{eq:flow}, seems to be
similar to that of adhering fibroblasts
\cite{fernandez2, monolayer}. At slow strain rates,
forces are low; above a critical strain rate, 
friction increases and the force becomes much stronger.
From our results, writing the critical rate as a strain we get
$\dot{V}_0/(r_0 A_0)\sim 5\,10^{-3}{\rm s}^{-1}$. 
In single fibroblasts, this is indeed the order of magnitude of the 
critical strain rate where frictional forces increase \cite{fernandez2},
which in turn agrees with the timescale of active processes \cite{thoumine,our_review}.
This suggests that changes in cell shape --including swelling--
take place at the rate allowed by spontaneous
unbinding of cytoskeletal crosslinks, while
faster changes are slowed down by
friction between cytoskeletal elements.

{\bf RVD is osmotically driven.}
Our results indicate that mechanical tension does not
provide the driving force for volume recovery.
The pearling modulation vanishes well before the
volume recovers, indicating zero membrane tension during RVD.
Moreover, in presence of cytoskeleton-perturbing drugs 
the volume recovery time $\tau$ does not increase, but 
becomes slightly shorter. This indicates that
volume recovery takes place via extrusion
of osmolites \cite{uruguayos}. Indeed, our simple model based on 
two ionic species suffices to explain the essential features
of the volume evolution curves. It captures
the broad response and even
the tendency to ``undershoot'' at large dilutions.

{\bf RVD follows an Arrhenius trend as a function of temperature.}
The timescales involved in the volume responses 
are temperature dependent. The relaxation time
is about an order of magnitude more temperature 
dependent than the swelling/shrinking rate, showing an exponential
reduction as the temperature is increased. Thus the ability
of the cell to respond to and negate perturbations to its
volume improves drastically as the temperature approaches
the normal physiological values. Due to this reason the 
neurites are able to withstand strong hypoosmotic-shocks at
higher temperatures when the same shock would have made
them rupture at a slightly lower temperature.

{\bf A dynamic picture of stretch injury.}
Similar shape transformations have been observed in nerves under the
name of ``beading'' as a response to stretch injury \citep{ochs1,ochs2}.
Interestingly, electron microscopy observations of the ultrastructure of stretch-beaded nerves
show that microtubules are splayed out in the beads \citep{ochs4},
consistent with our picture of a mechanical connection between
microtubules and the membrane.
The shapes of beaded nerves have been interpreted as equilibrium
shapes with a constant curvature \citep{ochs1};
however, our results offer an alternative explanation,
namely that the shape is given by the fastest growing mode
at the time of the increase in tension. If so, the shape
of beaded nerves would not simply follow from structural
properties (as expected for an equilibrium shape) but
would also be defined by the precise way stretch-injury
takes place, i.e., the rate and extent of loading.
Future studies may address this question in detail.

\section*{Conclusions}

Neurites respond
to sudden osmotic pressure changes
with a fast volume regulation response. 
The initial phase is characterised by a
nonlinear dependence of swelling rate
on the initial osmotic pressure.
Cytoskeletal perturbation, especially microtubule disruption,
accelerates swelling and increases the maximum volume
reached, but does not affect the relaxation phase.
Taking our results together, we propose
that mechanical forces due to the
nonlinear viscosity of the cytoskeleton
slow down the initial phase of change of volume.
This may provide instantaneous integrity to the neurite
while osmotic mechanisms ``warm up''.

\subsubsection*{Acknowledgements}

Experiments were performed at
the Universit\"at Bayreuth, Germany,
with the generous financial support of Albrecht Ott.
We thank Osvaldo Chara, Pablo Schwarzbaum, Karina Alleva and Wolfram Hartung for helpful discussions.

\bibliography{../../biblio/Pearling,../../../biblio2/cellpulling}

\begin{thebibliography}{54}
\providecommand{\url}[1]{\texttt{#1}}
\providecommand{\urlprefix}{ }

\bibitem[Weiss(1996)]{weiss}
Weiss, T., 1996.
\newblock Cellular Biophysics.
\newblock MIT Press.

\bibitem[Lang et~al.(1998)Lang, Busch, Ritter, V\"olkl, Waldegger, Gulbins, and
  H\"aussinger]{florian}
Lang, F., G.~Busch, M.~Ritter, H.~V\"olkl, S.~Waldegger, E.~Gulbins, and
  D.~H\"aussinger, 1998.
\newblock Functional significance of cell volume regulatory mechanisms.
\newblock \emph{Physiol. Reviews} 78:247.

\bibitem[Hernandez and Cristina(1998)]{uruguayos}
Hernandez, J., and E.~Cristina, 1998.
\newblock {Modelling cell volume regulation in nonexcitable cells: the roles of
  the Na$^+$ pump and of cotransport systems}.
\newblock \emph{Am. Physiol. Soc.} C1067.

\bibitem[MacKnight(1987)]{macknight}
MacKnight, A. D.~C., 1987.
\newblock Volume Maintenance in Isosmotic Conditions.
\newblock \emph{In} R.~Gilles, A.~Kleinzeller, and L.~Bolis, editors, Cell
  Volume Control: Fundamental and comparative aspects in animal cells. Academic
  Press, San Diego, 3--43.

\bibitem[Bray(2001)]{braybook}
Bray, D., 2001.
\newblock {Cell Movements: from molecules to motility}.
\newblock Garland Publishing, Inc., New York, 2nd edition.

\bibitem[Alberts et~al.(1994)Alberts, Bray, Lewis, Raff, Roberts, and
  Watson]{alberts}
Alberts, B., D.~Bray, J.~Lewis, M.~Raff, K.~Roberts, and J.~D. Watson, 1994.
\newblock {Molecular Biology of the Cell}.
\newblock Garland Publishing, Inc., New York, 3rd edition.

\bibitem[Janmey(1998)]{cytoskel_signal_janmey}
Janmey, P.~A., 1998.
\newblock The cytoskeleton and cell signaling: component localization and
  mechanical coupling.
\newblock \emph{Physiol. Rev.} 78:763--781.

\bibitem[Fabry et~al.(2001)Fabry, Maksym, Butler, Glogauer, Navajas, and
  Fredberg]{glassy1}
Fabry, B., G.~N. Maksym, J.~P. Butler, M.~Glogauer, D.~Navajas, and J.~J.
  Fredberg, 2001.
\newblock Scaling the microrheology of living cells.
\newblock \emph{Phys. Rev. Lett.} 87:148102.

\bibitem[Thoumine and Ott(1997)]{thoumine}
Thoumine, O., and A.~Ott, 1997.
\newblock Time scale dependent viscoelastic and contractile regimes in
  fibroblasts probed by microplate manipulation.
\newblock \emph{J. Cell. Sci.} 110:2109--2116.

\bibitem[Fern\'andez et~al.(2006)Fern\'andez, Pullarkat, and Ott]{mipaper}
Fern\'andez, P., P.~A. Pullarkat, and A.~Ott, 2006.
\newblock A master relation defines the nonlinear viscoelasticity of single
  fibroblasts.
\newblock \emph{Biophys. J.} 90:3796--3805.

\bibitem[Pullarkat et~al.(2007)Pullarkat, Fern\'andez, and Ott]{our_review}
Pullarkat, P.~A., P.~A. Fern\'andez, and A.~Ott, 2007.
\newblock Rheological properties of the Eukaryotic cell cytoskeleton.
\newblock \emph{Phys. Rep.} 449:29--53.

\bibitem[Bernal et~al.(2007)Bernal, Pullarkat, and Melo]{pramod07_axon-mech}
Bernal, R., P.~A. Pullarkat, and F.~Melo, 2007.
\newblock Mechanical properties of axons.
\newblock \emph{Phys. Rev. Lett.} 99:018301.

\bibitem[Henson(1999)]{henson}
Henson, J., 1999.
\newblock Relationships between the Actin Cytoskeleton and Cell Volume
  Regulation.
\newblock \emph{Microscopy research and technique} 47:155--162.

\bibitem[Kleinzeller(1965)]{kleinzeller}
Kleinzeller, A., 1965.
\newblock The volume regulation in some animal cells.
\newblock \emph{Arch. Biol.} 76:217--232.

\bibitem[Cantielo(1997)]{cantiello}
Cantielo, H., 1997.
\newblock Role of actin filament organization in cell volume and ion channel
  regulation.
\newblock \emph{J. Exp. Zoology} 279:425.

\bibitem[Mills(1987)]{mills}
Mills, J., 1987.
\newblock The Cell Cytoskeleton: Possible Role in Volume Control.
\newblock \emph{In} R.~Gilles, A.~Kleinzeller, and L.~Bolis, editors, Cell
  volume control: fundamental and comparative aspects in animal cells. Academic
  Press,Inc., San Diego, 75--101.

\bibitem[Strieter et~al.(1990)Strieter, Stephenson, Palmer, and
  Weinstein]{strieter}
Strieter, J., J.~L. Stephenson, L.~G. Palmer, and A.~M. Weinstein, 1990.
\newblock Volume-activated chloride permeability can mediate cell volume
  regulation in a mathematical model of a tight epithelium.
\newblock \emph{J. Gen. Physiol.} 96:319--344.

\bibitem[Strange(2004)]{volume_strange}
Strange, K., 2004.
\newblock Cellular volume homeostasis.
\newblock \emph{Adv.Physiol.Educ.} 28:155--159.

\bibitem[Downey et~al.(1995)Downey, Grinstein, Sue-A-Quan, Czaban, and
  Chan]{downey}
Downey, G., S.~Grinstein, A.~Sue-A-Quan, B.~Czaban, and C.~Chan, 1995.
\newblock Volume Regulation in leukocytes: requirement for an intact
  cytoskeleton.
\newblock \emph{J. Cell Physiol.} 163:96--104.

\bibitem[Pullarkat et~al.(2006)Pullarkat, Dommersnes, Fern\'andez, Joanny, and
  Ott]{pramods}
Pullarkat, P., P.~Dommersnes, P.~Fern\'andez, J.-F. Joanny, and A.~Ott, 2006.
\newblock Osmotically induced shape transformations in axons.
\newblock \emph{Phys. Rev. Lett.} 96:048104.

\bibitem[Wan et~al.(1995)Wan, Harris, and Morris]{neuronsextreme}
Wan, X., J.~Harris, and C.~Morris, 1995.
\newblock Responses of neurons to extreme osmomechanical stress.
\newblock \emph{J. Memb. Biol.} 145:21--31.

\bibitem[Heubusch et~al.(1985)Heubusch, Jung, and
  Green]{osmo_erythro_cytoskel_heubusch}
Heubusch, P., C.~Jung, and F.~Green, 1985.
\newblock The osmotic response of human erythrocytes and the membrane
  cytoskeleton.
\newblock \emph{J.Cell.Physiol} 122:266--272.

\bibitem[Suchyna et~al.(2004)Suchyna, Besch, and Sachs]{suchyna}
Suchyna, T., S.~Besch, and F.~Sachs, 2004.
\newblock Dynamic regulation of mechanosensitive channels: capacitance used to
  monitor patch tension in real time.
\newblock \emph{Phys Biol} 1:1.

\bibitem[D'Alessandro et~al.(2002)D'Alessandro, Russell, Morley, Davies, and
  Birgitte~Lane]{d'alessandro}
D'Alessandro, M., D.~Russell, S.~M. Morley, A.~M. Davies, and E.~Birgitte~Lane,
  2002.
\newblock Keratin mutations of epidermolysis bullosa simplex alter the kinetics
  of stress response to osmotic shock.
\newblock \emph{J. Cell Sci.} 115:4341--4351.

\bibitem[Cornet et~al.(1988)Cornet, Delpire, and Gilles]{cornet88}
Cornet, M., E.~Delpire, and R.~Gilles, 1988.
\newblock Relations between cell volume control, microfilaments and
  microtubules networks in T2 and PC12 cultured cells.
\newblock \emph{J. Physiol. Paris} 83:43--49.

\bibitem[Light et~al.(2003)Light, Attwood, Siegel, and
  Baumann]{Ca_swelling_Light}
Light, D., A.~Attwood, C.~Siegel, and N.~Baumann, 2003.
\newblock Cell swelling increases intracellular calcium in {\it Necturus}
  erythrocytes.
\newblock \emph{J.Cell Sci.} 116:101--109.

\bibitem[Arora et~al.(1994)Arora, Bibby, and
  McCulloch]{oscillations_Ca_stretch}
Arora, P., K.~Bibby, and C.~McCulloch, 1994.
\newblock Slow oscillations of free intracellular calcium ion concentration in
  human fibroblasts responding to mechanical stretch.
\newblock \emph{J.Cell.Physiol.} 161:187--200.

\bibitem[Guilak et~al.(1999)Guilak, Zell, Erickson, Grande, Rubin, McLeod, and
  Donahue]{Ca-waves_Ga}
Guilak, F., R.~Zell, G.~Erickson, D.~Grande, C.~Rubin, K.~McLeod, and
  H.~Donahue, 1999.
\newblock Mechanically induced calcium waves in articular chondrocytes are
  inhibited by gadolinium and amiloride.
\newblock \emph{J.Orthop.Res.} 17:421--429.

\bibitem[Greene and Tischler(1976)]{PC12_establishment}
Greene, L., and A.~Tischler, 1976.
\newblock Establishment of a noradrenergic clonal line of rat adrenal
  pheochromocytoma cells which respond to nerve growth factor.
\newblock \emph{Proc.Natl.Acad.Sci.USA} 73:2424--2428.

\bibitem[Roger~Jacobs and Stevens(1986)]{changes_neurite_cytoskel_jacobs}
Roger~Jacobs, J., and J.~Stevens, 1986.
\newblock Changes in the organization of the neuritic cytoskeleton during nerve
  growth factor-activated differentiation of PC12 cells: a serial electron
  microscopic study of the development and control of neurite shape.
\newblock \emph{J.Cell Biol.} 103:895--906.

\bibitem[Markin et~al.(1999)Markin, Tanelian, Jersild, and Ochs]{ochs1}
Markin, V., D.~Tanelian, R.~Jersild, Jr., and S.~Ochs, 1999.
\newblock Biomechanics of stretch-induced beading.
\newblock \emph{Biophys.J.} 76:2852--2860.

\bibitem[Ochs et~al.(1996)Ochs, Pourmand, and Jersild]{ochs2}
Ochs, S., R.~Pourmand, and R.~Jersild, Jr., 1996.
\newblock Origin of beading constrictions at the axolemma: presence in
  unmyelinated axons and after $\beta$,$\beta^\prime$-iminodipropionitrile
  degradation of the cytoskeleton.
\newblock \emph{Neuroscience} 70:1081--1096.

\bibitem[Drexler et~al.(2001)Drexler, Dirks, MacLeod, Quentmeier, Steube, and
  Uphoff]{DSMZ}
Drexler, H.~G., W.~Dirks, W.~R. A.~F. MacLeod, H.~Quentmeier, K.~G. Steube, and
  C.~C. Uphoff, 2001.
\newblock DSMZ Catalogue of Human and Animal Cell Lines.
\newblock Braunschweig.

\bibitem[Greene et~al.(1991)Greene, Sobeih, and Teng]{PC12_culture_greene}
Greene, L., M.~Sobeih, and K.~Teng, 1991.
\newblock Methodologiesfor the culture and experimental use of the PC12 rat
  pheochromocytoma cell line.
\newblock \emph{In} G.~Banker, and K.~Goslin, editors, Culturing Nerve Cells.
  MIT Press, Cambridge, 207--226.

\bibitem[De~Brabander et~al.(1976)De~Brabander, Van~de Veire, Aerts, Borgers,
  and Janssen]{nocodazol}
De~Brabander, M., R.~Van~de Veire, R.~Aerts, M.~Borgers, and P.~Janssen, 1976.
\newblock {The effects of
  methyl[5-(2-thienylcarbonyl)-1H-benzimidazol-2-yl]carbamate (R 17 934, NSC
  238159), a new synthetic antitumoral drug interfering with microtubules, on
  mammalian cells cultured in vitro}.
\newblock \emph{Cancer Res.} 36:905--916.

\bibitem[Straight et~al.(2003)Straight, Cheung, Limouze, Chen, Westwood,
  Sellers, and Mitchison]{blebbistatin}
Straight, A.~F., A.~Cheung, J.~Limouze, I.~Chen, N.~J. Westwood, J.~R. Sellers,
  and T.~J. Mitchison, 2003.
\newblock {Dissecting temporal and spatial control of cytokinesis with a Myosin
  II inhibitor}.
\newblock \emph{Science} 299:1743--1747.

\bibitem[van Hoek et~al.(1990)van Hoek, de~Jong, and van Os]{dmso_perm_hoek}
van Hoek, A.~N., M.~D. de~Jong, and C.~H. van Os, 1990.
\newblock Effects of dimethylsulfoxide and mercurial sulfhydryl reagents on
  water and solute permeability of rat kidney brush border membranes.
\newblock \emph{Biochim.Biophys.Acta} 1030:203--210.

\bibitem[Maric et~al.(2001)Maric, Wiesner, Lorenz, Klussmann, Betz, and
  Rosenthal]{Wperm_renal_maric}
Maric, K., B.~Wiesner, D.~Lorenz, E.~Klussmann, T.~Betz, and W.~Rosenthal,
  2001.
\newblock Cell volume kinetics of adherent epithelial cells measured by laser
  scanning reflection microscopy: determination of water permeability changes
  of renal principal cells.
\newblock \emph{Biophys.J.} 80:1783--1790.

\bibitem[Huster et~al.(1997)Huster, Jin, Arnold, and
  Gawrisch]{Wperm_memb_huster}
Huster, D., A.~Jin, K.~Arnold, and K.~Gawrisch, 1997.
\newblock {Water permeability of polyunsaturated lipid membranes measured by
  ${}^{17}$O NMR}.
\newblock \emph{Biophys.J.} 73:855--864.

\bibitem[Lucio et~al.(2003)Lucio, Santos, and Mesquita]{brasileros}
Lucio, A., R.~Santos, and O.~Mesquita, 2003.
\newblock Measurements and modeling of water transport and osmoregulation in a
  single kidney cell using optical tweezers and videomicroscopy.
\newblock \emph{Phys. Rev. E} 68:041906.

\bibitem[Fern\'andez(2006)]{mitesis}
Fern\'andez, P., 2006.
\newblock Mechanics of living cells: nonlinear viscoelasticity of single
  fibroblasts and shape instabilities in axons.
\newblock Ph.D. thesis, Universit\"at Bayreuth.

\bibitem[Blumenfeld and Tikhonov(1994)]{biophysics_blumenfeld}
Blumenfeld, L.~A., and A.~N. Tikhonov, 1994.
\newblock Biophysical thermodynamics of intracellular processes: molecular
  machines of the living cell.
\newblock Springer, Berlin.

\bibitem[D.~Kilinc and Barbee(2008)]{08barbee_beading}
D.~Kilinc, G.~G., and K.~A. Barbee, 2008.
\newblock Mechanicall-induced membrane portaion causes axonal beading and
  localized cytoskeletal damage.
\newblock \emph{Exp. Neurol.} 212:422--430.

\bibitem[Ochs et~al.(1994)Ochs, Jersild, Pourmand, and Potter]{ochs4}
Ochs, S., R.~Jersild, Jr., R.~Pourmand, and C.~Potter, 1994.
\newblock The beaded form of myelinated nerve fibers.
\newblock \emph{Neuroscience} 61:361--372.

\bibitem[Bar-Ziv and Moses(1994)]{barziv1}
Bar-Ziv, R., and E.~Moses, 1994.
\newblock Instability and ``pearling'' states produced in tubular membranes by
  competition of curvature and tension.
\newblock \emph{Phys. Rev. Lett.} 73:1392--1395.

\bibitem[Bar-Ziv et~al.(1998)Bar-Ziv, Moses, and Nelson]{barziv3}
Bar-Ziv, R., E.~Moses, and P.~Nelson, 1998.
\newblock Dynamic excitations in membranes induced by optical tweezers.
\newblock \emph{Biophys.J.} 75:294--320.

\bibitem[Bernal et~al.(2009)Bernal, Melo, and Pullarkat]{pramod09_drag}
Bernal, R., F.~Melo, and P.~A. Pullarkat, 2009.
\newblock Drag force as a tool to test the active mechanical response of PC12
  neurites.
\newblock \emph{in press in Biophys. J.} .

\bibitem[Bar-Ziv and Moses(1999)]{barziv2}
Bar-Ziv, R., and E.~Moses, 1999.
\newblock Pearling in cells: a clue to understanding cell shape.
\newblock \emph{Proc. Natl. Acad. Sci. USA} 96:10140--10145.

\bibitem[Nelson et~al.(1995)Nelson, Powers, and Seifert]{pearling1}
Nelson, P., T.~Powers, and U.~Seifert, 1995.
\newblock Dynamic theory of pearling instability in cylindrical vesicles.
\newblock \emph{Phys. Rev. Lett.} 74:3384.

\bibitem[Rayleigh(1892)]{rayleigh}
Rayleigh, L., 1892.
\newblock On the instability of a cylinder of viscous liquid under capillary
  force.
\newblock \emph{Philos.Mag.} 34:195.

\bibitem[Chandrasekhar(1970)]{chandrasekhar}
Chandrasekhar, S., 1970.
\newblock Hydrodynamic and hydromagnetic stability.
\newblock Dover, New York, 3 edition.

\bibitem[Fern\'andez and Ott(2008)]{fernandez2}
Fern\'andez, P., and A.~Ott, 2008.
\newblock Single cell mechanics: stress stiffening and kinematic hardening.
\newblock \emph{Phys. Rev. Lett.} 100:238102.

\bibitem[Cornet et~al.(1993)Cornet, Ubl, and Kolb]{cornet93}
Cornet, M., J.~Ubl, and H.-A. Kolb, 1993.
\newblock Cytoskeleton and ion movements during Volume regulation in cultured
  PC12 cells.
\newblock \emph{J. Membrane Biol.} 133:161--170.

\bibitem[Fern\'andez et~al.(2007)Fern\'andez, Heymann, Ott, Aksel, and
  Pullarkat]{monolayer}
Fern\'andez, P., L.~Heymann, A.~Ott, N.~Aksel, and P.~A. Pullarkat, 2007.
\newblock Shear rheology of a cell monolayer.
\newblock \emph{New J. Phys.} 9:419.

\end{thebibliography}

\end{document}